\documentclass[article,nojss,shortnames]{jss}\usepackage[]{graphicx}\usepackage[]{color}
\makeatletter
\def\maxwidth{ %
  \ifdim\Gin@nat@width>\linewidth
    \linewidth
  \else
    \Gin@nat@width
  \fi
}
\makeatother

\usepackage{Sweave}

\author{
Antoine Filipovi\'c-Pierucci\\URC-Eco
\And
Kevin Zarca\\URC-Eco
\AND
Isabelle Durand-Zaleski\\URC-Eco
}
\title{Markov Models for Health Economic Evaluations: The \proglang{R} Package \pkg{heemod}}

\Plainauthor{Antoine Filipovi\'c-Pierucci, Kevin Zarca, Isabelle Durand-Zaleski} 
\Plaintitle{Markov Models for Health Economic Evaluation: The R Package heemod} 
\Shorttitle{\pkg{heemod}: Models For Health Economic Evaluation in \proglang{R}} 

\Abstract{
Health economic evaluation studies are widely used in public health to assess health strategies in terms of their cost-effectiveness and inform public policies. We developed an \proglang{R} package for writing Markov models for health economic evaluations which implements the modelling and reporting features described in reference textbooks and guidelines: deterministic and probabilistic sensitivity analysis, heterogeneity analysis, time dependency on state-time and model-time (semi-Markov and non-homogeneous Markov models), etc. In this paper we illustrate the features of \pkg{heemod} by building and analysing an example Markov model. We then explain the design and the underlying implementation of the package.
}
\Plainkeywords{health economic evaluation, Markov models, R} 

\Address{
  Antoine Filipovi\'c-Pierucci\\
  Unit\'e de recherche clinique en \'economie de la sant\'e d'\^Ile-de-France\\
  H\^otel-Dieu\\
  Assistance publique-H\^opitaux de Paris\\
  Parvis de Notre-Dame, Paris, France\\
  E-mail: \email{antoine.filipovic-pierucci@urc-eco.fr}\\
  URL: \url{https://pierucci.org}
}

\usepackage{mathtools}
\usepackage{multicol}

\IfFileExists{upquote.sty}{\usepackage{upquote}}{}
\begin{document}


\section{Introduction} \label{sec:intro}

Health economic evaluation studies are widely used in public health to assess healthcare strategies in terms of their cost-effectiveness and inform public policies~\citep{russel1996}. In order to account for the long-term consequences of healthcare strategies, models are needed to extrapolate results to a longer time frame~\citep{sonnenberg1993, eddy2012}.\footnote{Usually the entire target population lifetime.}

These models estimate the repartition of a population in various health states (e.g. healthy, sick, dead) and how its health is affected by different strategies. Costs (e.g. medical or drug costs) and outcomes (e.g. life years or quality of life) are attached to each distinct health status, allowing to estimate the cost and effectiveness expected for every studied strategy.

By representing health status as states and health changes over time as transitions probabilities between states this process can be modelled with Markov chains, using a Markov model. Transition probabilities between states can be described by a square 2-dimensional transition matrix $T$, where element $i,j$ is the transition probability between state $i$ and $j$. The probability of being in a given state at time $t$ is given by:
\begin{equation} \label{eq:mat_2}
X \times T^t
\end{equation}
Where $X$ is a vector\footnote{Of length equal to the number of states.} giving the probability of being in a given state at the start of the model, and $T^t$ is the product of multiplying $t$ matrices $T$. The use of Markov models in health economic evaluation have been thoroughly described in \citet{beck1983}, \citet{sonnenberg1993} and \citet{briggs1998}.

In order to best inform the decision process, Markov models should incorporate a wide range of information to account for all the available evidence at a given time~\citep{briggs1998}. Results from various sources can be combined, such as estimated drug efficacy from clinical trials, disease evolution from epidemiological cohorts, quality of life values from population-level studies, transition probabilities from life-tables, etc. An implementation of Markov models should be flexible enough to receive all these sources of data.

Most Markov models are built using basic spreadsheet software such as \proglang{Microsoft Excel}~\citep{excel} or commercial packages such as \proglang{TreeAge}~\citep{treeage}, which has drawbacks: analyses are hard to reproduce and lack transparency, errors are difficult to spot, track and correct, and graphic capabilities are lacking~\citep{williams2016}. The \proglang{R} language~\citep{r} can overcome these issues through script-based approaches: there is a written record of what was done, the calculations are transparent, modification can be easily applied to the model, traceability is guaranteed,\footnote{Especially with the help of version control software such as \proglang{git}~\citep{git} that are particularly well suited to script files.} and the code just needs to be run again to reproduce the analysis~\citep{williams2016}. Despite these advantages, usage of \proglang{R} in health economic modelling has been limited by the significant challenge of programming Markov models from scratch and the lack of packages providing a comprehensive set of tools for developing such models.

Our objective was to develop an \proglang{R} package for writing Markov models for health economic evaluations, using a simple declarative syntax, which implements the modelling and reporting features described in reference textbooks~\citep{drummond2005,briggs2006} and guidelines~\citep{eddy2012,husereau2013}. We named the package \pkg{heemod}, standing for Health Economic Evaluation MODelling. The package is available from the Comprehensive R Archive Network (CRAN) at \url{http://CRAN.R-project.org/package=heemod}.

In Section~\ref{sec:example} we illustrate the possibilities of \pkg{heemod} by building and analysing an example Markov model. For completeness we then present in Section~\ref{sec:features} the features that were not used in the previous example. In Section~\ref{sec:math} we detail the mathematical implementation of some functionalities. Finally in Section~\ref{sec:impl} we explain the design and the back-end of the package.

\section{Building and analysing a model: an example} \label{sec:example}

In this section we use a simplified example to illustrate how and why Markov models are used in health economic evaluation studies. We introduce theoretical concepts and methods along as they are encountered, present the production of results with the \pkg{heemod} package, and their interpretation. In this article we used \pkg{heemod} version 0.9.0.9001.

\subsection{Description of the question}

For this example we will model the imaginary disease called \emph{shame}, which is still a terminal disease in some parts of the Galaxy~\citep{adams1979}. At the onset the disease is asymptomatic: patients are quite unashamed, they do not feel sick, have a good quality of life, and are not likely to die of shame. But patients are at risk of being ashamed: that marks the entry into the symptomatic phase of the disease, with frequent hospital stays, deteriorated quality of life and a high risk of dying of shame. The probability to revert to the asymptomatic unashamed state is unfortunately quite low, and there is no cure known to work reliably: to be effective shame therapy should thus be provided during the initial asymptomatic state, to prevent being ashamed in the first place.

\subsection{Compared strategies}

Three strategies are proposed to prevent being ashamed:
\begin{description}
  \item [Base strategy (\code{base}):] Do nothing, this is the natural evolution of the disease.
  \item [Medical treatment (\code{med}):] Patients with asymptomatic disease are treated with a \emph{ashaminib}, a highly-potent shame inhibitor, until progression to symptomatic state in order to lower the risk of being ashamed.
  \item [Surgical treatment (\code{surg}):] Patients with asymptomatic disease undergo \emph{shamectomy}, a surgical procedure that lowers the risk of being ashamed. The procedure needs to be performed only once.
\end{description}

Medical treatment is effective in preventing symptomatic disease, but the drug is expensive. Surgical treatment is a one-time cost, but its effect decreases with time. The increased probability of dying of shame once in the symptomatic disease state does not depend on the treatment used before, when the disease was asymptomatic.

\subsection{States}

From the description of the disease we can define 3 states:
\begin{description}
  \item [Asymptomatic state (\code{pre}):] Before the symptomatic state, when treatment can still be provided.
  \item [Symptomatic state (\code{symp}):] Symptomatic disease, after being ashamed. With degraded health, high hospital costs and increased probability of dying of shame.
  \item [Death (\code{death}):] Death by natural causes or because of shame.
\end{description}

\subsection{Model parameters}

In this section we define parameters that will be called later in the analysis (e.g. in the transition matrix or the state values). Because we said in the disease description that some probabilities and values vary with time, we need to introduce some concepts regarding time-dependency before we can define the parameters.

Transition probabilities or state values may change with time (e.g. the protecting effect of surgery may decrease with time after the procedure, probability of all-causes death may increase as the population gets older, hospital costs may change with disease evolution). It is thus important to account for time-dependency in order to build accurate models. In Markov models values may depend on 2 distinct measurements of time~\citep{hawkins2005}: time elapsed since the start of the model (called \emph{model time}), and time spent in a given state (called \emph{state time}). Both situations can co-exist in a same model.

In \pkg{heemod} time-dependency is specified with 2 variables: \code{model_time}\footnote{Or its alias \code{markov\_cycle}.} and \code{state_time}. These package-reserved names return sequential values starting from 1, corresponding to time spent in the model for \emph{model time} and time spent in a given state for \emph{state time}. They can be used in any user-defined expression or function.

In our case the probability of all-cause death depends on age. Because age increases with time spent since the beginning of the model, the all-cause death probability is \emph{model time} dependent. On the other hand the probability of dying of shame depends on the time elapsed after being ashamed (i.e. time spent in the \code{symp} state): this probability is \emph{state time} dependent. The probability of being ashamed after surgery, the cost of surgery, and the hospital costs in the symptomatic state also depend on the time spent in their respective state.

In this model we will use a cycle duration of 1 year: we must take care that all transitions probabilities, values attached to states, and discount rates are calculated on this time-frame.

We can now create the global parameters with \code{define_parameters()}:
\begin{Schunk}
\begin{Sinput}
R> par_mod <- define_parameters(
R+   age_base  = 20,
R+   age_cycle = model_time + age_base)
\end{Sinput}
\end{Schunk}
The age of individuals for a given cycle \code{age_cycle} is the age at the beginning of the model (\code{age_base}), plus the time the model has run (\code{model_time}).
\begin{Schunk}
\begin{Sinput}
R> par_mod <- modify(
R+   par_mod,
R+   
R+   sex_indiv   = "MLE", # MLE => male in the WHO database
R+   p_death_all = get_who_mr(
R+     age     = age_cycle,
R+     sex     = sex_indiv,
R+     country = "GBR",
R+     local   = TRUE))
\end{Sinput}
\end{Schunk}
The death probability \code{p_death_all}, as a function of age and sex, is fetched from the World Health Organisation database with \code{get_who_mr()},\footnote{Relying on the \pkg{rgho} package~\citep{rgho}.} here for a British population.\footnote{We specify the use of local data cached in \pkg{heemod} with \code{local = TRUE} to avoid adding overhead query time.}
\begin{Schunk}
\begin{Sinput}
R> par_mod <- modify(
R+   par_mod,
R+   
R+   p_death_disease = compute_surv(
R+     fit_death_disease,
R+     time     = state_time,
R+     km_limit = 5))
\end{Sinput}
\end{Schunk}
The probability of dying of shame when the disease is symptomatic \code{p_death_disease} is extracted with the \code{get_probs_from_surv()} function from \code{fit_death_disease}, a model fitted with the \pkg{flexsurv} package~\citep{flexsurv}. Because this probability depends on time spent in the \code{disease} state, the \code{state_time} model variable is used to specify time. Here we use non-parametric Kaplan-Meier estimates for the first 5 years instead of model-fitted values with \code{km_limit = 5}.

The parametric survival model \code{fit_death_disease} used to compute \code{p_death_disease} is fitted with the following code:
\begin{Schunk}
\begin{Sinput}
R> fit_death_disease <- flexsurv::flexsurvreg(
R+   survival::Surv(time, status) ~ 1,
R+   dist = "weibull",
R+   data = tab_surv)
\end{Sinput}
\end{Schunk}
Where \code{tab_surv} is a data-frame containing survival data.
\begin{Schunk}
\begin{Sinput}
R> dput(tab_surv)
\end{Sinput}
\begin{Soutput}
structure(list(time = c(0.4, 8.7, 7, 5.1, 9.2, 1, 0.5, 3.3, 1.8, 
3, 6.7, 3.7, 1.1, 5.9, 5.1, 10, 10, 10, 10, 10, 10, 10, 10, 10, 
10), status = c(1L, 1L, 1L, 1L, 1L, 1L, 1L, 1L, 1L, 1L, 1L, 1L, 
1L, 1L, 1L, 0L, 0L, 0L, 0L, 0L, 0L, 0L, 0L, 0L, 0L)), .Names = c("time", 
"status"), row.names = c(NA, -25L), class = "data.frame")
\end{Soutput}
\end{Schunk}
\begin{Schunk}
\begin{Sinput}
R> par_mod <- modify(
R+   par_mod,
R+   
R+   p_death_symp = combine_probs(
R+     p_death_all,
R+     p_death_disease))
\end{Sinput}
\end{Schunk}
The death probability in the symptomatic state \code{p_death_symp} is the probability to die either from old age (\code{p_death_all}) or from the disease (\code{p_death_symp}). Assuming those probabilities are independent, we use the \code{combine_probs()} to combine them with the formula $P(A \cup B)=1-\big(1-P(A)\big)\times \big(1-P(B)\big)$.
\begin{Schunk}
\begin{Sinput}
R> par_mod <- modify(
R+   par_mod,
R+   
R+   p_disease_base = 0.25,
R+   med_effect     = 0.5,
R+   p_disease_med  = p_disease_base * med_effect)
\end{Sinput}
\end{Schunk}
The probability of disease under medical treatment \code{p_disease_med} is the base probability of disease \code{p_disease_base} times the protecting effect of the treatment, \code{med_effect}.
\begin{Schunk}
\begin{Sinput}
R> par_mod <- modify(
R+   par_mod,
R+   
R+   shape = 1.5, # We will see later why we need
R+   scale = 5,   # to define these 2 parameters here.
R+   p_disease_surg = define_survival(
R+       distribution = "weibull",
R+       shape        = shape,
R+       scale        = scale) %>% 
R+     compute_surv(time = state_time))
\end{Sinput}
\end{Schunk}
The probability of disease after surgery is extracted with \code{get_probs_from_surv()} from a parametric Weibull survival model defined with \code{define_survival()}. For reason explained in Section~\ref{sec:psa} parameters \code{scale} and \code{shape} are not written in the \code{define_survival()} call, but defined separately.
\begin{Schunk}
\begin{Sinput}
R> par_mod <- modify(
R+   par_mod,
R+   
R+   cost_surg       = 20000,
R+   cost_surg_cycle = ifelse(state_time == 1, cost_surg, 0))
\end{Sinput}
\end{Schunk}
Because surgery is only performed once at the beginning of the \code{pre} state, the time-dependant variable \code{state_time} was used to limit surgery costs to the first cycle in the \code{pre} state.
\begin{Schunk}
\begin{Sinput}
R> par_mod <- modify(
R+   par_mod,
R+   
R+   cost_hospit_start = 11000,
R+   cost_hospit_end   = 9000,
R+   n_years           = 9,
R+   cost_hospit_cycle = ifelse(
R+     state_time < n_years,
R+     cost_hospit_start,
R+     cost_hospit_end))
\end{Sinput}
\end{Schunk}
After \code{n_years} in the symptomatic state the symptoms become milder and hospital costs decrease (from \code{cost_hospit_start} to \code{cost_hospit_end}). We used the time-dependant variable \code{state_time} to condition the hospital costs \code{cost_hospit} on \code{n_years}.
\begin{Schunk}
\begin{Sinput}
R> par_mod <- modify(
R+   par_mod,
R+   
R+   p_cured      = 0.001,
R+   cost_med     = 5000,
R+   dr           = 0.05,
R+   qaly_disease = 0.5)
\end{Sinput}
\end{Schunk}
Finally we define \code{p_cured} the probability to spontaneously revert to the asymptomatic unashamed state, \code{cost_med} the drug costs, \code{dr} the discount rate for a year and \code{qaly_disease} the QALY for one year in the symptomatic state.

\subsection{Transitions}

\begin{figure}
\centering
\begin{Schunk}

\includegraphics[width=0.5\linewidth]{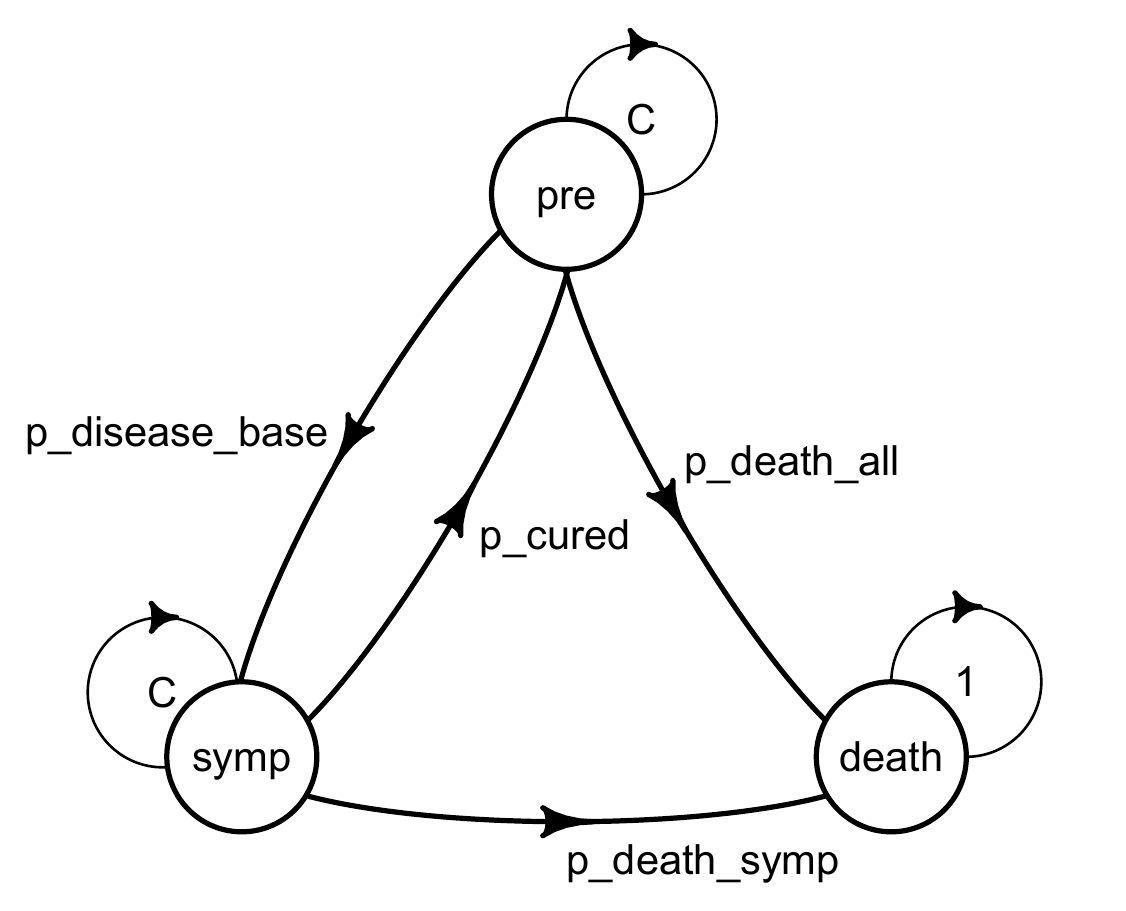} \end{Schunk}
\caption{Transition diagram for the \code{base} strategy.}
\label{fig:tm_base}
\end{figure}
We define a transition matrix for the \code{base} strategy with the \code{define_transition()} function. We can reference parameters defined in the previous section.
\begin{Schunk}
\begin{Sinput}
R> mat_base <- define_transition(
R+   state_names = c("pre", "symp", "death"),
R+   
R+   C,       p_disease_base, p_death_all,
R+   p_cured, C,              p_death_symp,
R+   0,       0,              1)
\end{Sinput}
\end{Schunk}
\begin{Schunk}
\begin{Sinput}
R> mat_base
\end{Sinput}
\begin{Soutput}
A transition matrix, 3 states.

      pre     symp           death       
pre   C       p_disease_base p_death_all 
symp  p_cured C              p_death_symp
death                        1           
\end{Soutput}
\end{Schunk}
\code{p_disease_base} is the probability of being ashamed in the \code{base} strategy, \code{p_death_all} the all cause probability of death (not caused by the disease) and \code{p_death_symp} the death probability in the symptomatic state (greater than \code{p_death_all}). \code{p_cured} is the unlikely probability to revert to the asymptomatic unashamed state. \code{p_cured}, \code{p_death_symp} and \code{p_death_all} do not depend on the strategy. The value of these parameters will be defined later. \code{C} is an alias for the probability complement, 1 minus the sum of probabilities in a given row. Death was modelled as an absorbing health state (i.e. the probability of transitioning from death to other health states was set to zero). The resulting transition diagram is presented in Figure~\ref{fig:tm_base}.\footnote{Generated by \code{plot(mat\_base)}.}

Similarly, transitions can be defined for the other 2 strategies. In our case only the name of the probabilities change: \code{p_disease_base} becomes \code{p_disease_med} or \code{p_disease_surg} (those parameters will also be defined later).
\begin{Schunk}
\begin{Sinput}
R> mat_med <- define_transition(
R+   state_names = c("pre", "symp", "death"),
R+   
R+   C,       p_disease_med, p_death_all,
R+   p_cured, C,             p_death_symp,
R+   0,       0,             1)
\end{Sinput}
\end{Schunk}
\begin{Schunk}
\begin{Sinput}
R> mat_surg <- define_transition(
R+   state_names = c("pre", "symp", "death"),
R+   
R+   C,       p_disease_surg, p_death_all,
R+   p_cured, C,              p_death_symp,
R+   0,       0,              1)
\end{Sinput}
\end{Schunk}

\subsection{State values}

Next we define the values associated with states using the \code{define_state} function. An arbitrary number of values can be attached to a state, here we define: \code{cost_treat} the treatment cost (drug costs for the \code{med} strategy or surgery costs for the \code{surg} strategy, there is no treatment in the \code{base} strategy), \code{cost_hospit} the hospitalization costs, \code{cost_total} the total cost, and \code{qaly} the health-related quality-adjusted life years (QALY), where 1 stands for one year in perfect health and 0 stands for death~\citep{torrance1989}. In the following code we define the state \code{pre}:
\begin{Schunk}
\begin{Sinput}
R> state_pre <- define_state(
R+   cost_treat  = dispatch_strategy(
R+     base = 0, # no treatment => no treatment cost
R+     med  = cost_med,
R+     surg = cost_surg_cycle),
R+   cost_hospit = 0, # good health => no hospital expenses
R+   cost_total  = discount(cost_treat + cost_hospit, r = dr),
R+   qaly        = 1)
\end{Sinput}
\end{Schunk}
To dispatch the cost of treatment according to the strategy we used the \code{dispatch_strategy()} function, with arguments named as strategies (we will define the strategy names in Section~\ref{sec:run_model}). Another approach would have been to define 3 versions of \code{state_pre}, one per strategy, and in the next section use the corresponding version in each distinct \code{define_strategy()} call.

The total cost is discounted with the \code{discount()} function at a given rate (\code{dr}). The QALY attached to 1 year in this state are set to 1, corresponding to 1 year in perfect health. The variables \code{dr}, \code{cost_med} and \code{cost_surg_cycle} will be defined later.

The other 2 states are defined similarly:
\begin{Schunk}
\begin{Sinput}
R> state_symp <- define_state(
R+   cost_treat  = 0,
R+   cost_hospit = cost_hospit_cycle,
R+   cost_total  = discount(cost_treat + cost_hospit, r = dr),
R+   qaly        = qaly_disease)
\end{Sinput}
\end{Schunk}
\begin{Schunk}
\begin{Sinput}
R> state_death <- define_state(
R+   cost_treat  = 0,
R+   cost_hospit = 0,
R+   cost_total  = 0,
R+   qaly        = 0)
\end{Sinput}
\end{Schunk}
Patients have a degraded quality of life and are hospitalized during the symptomatic disease phase, we need to define specific QALYs (\code{qaly_disease}) and hospital costs (\code{cost_hospit_cycle}) for this state. These variables will be defined later. Finally dead patients have QALYs at 0, and they do not cost anything to the healthcare system.

\subsection{Strategies}

All the information (states and transitions) is now available to define the strategies. For this purpose we use the \code{define_strategy()} function. Only the transition objects differ between \code{strat_base}, \code{strat_med} and \code{strat_surg}.
\begin{Schunk}
\begin{Sinput}
R> strat_base <- define_strategy(
R+   transition = mat_base,
R+   
R+   pre   = state_pre,
R+   symp  = state_symp,
R+   death = state_death)
\end{Sinput}
\end{Schunk}
\begin{Schunk}
\begin{Sinput}
R> strat_med <- define_strategy(
R+   transition = mat_med,
R+   
R+   pre   = state_pre,
R+   symp  = state_symp,
R+   death = state_death)
\end{Sinput}
\end{Schunk}
\begin{Schunk}
\begin{Sinput}
R> strat_surg <- define_strategy(
R+   transition = mat_surg,
R+   
R+   pre   = state_pre,
R+   symp  = state_symp,
R+   death = state_death)
\end{Sinput}
\end{Schunk}

\subsection{Running the model} \label{sec:run_model}

The model can then be run with \code{run_model()}:
\begin{Schunk}
\begin{Sinput}
R> res_mod <- run_model(
R+   parameters = par_mod,
R+   
R+   base = strat_base,
R+   med  = strat_med,
R+   surg = strat_surg,
R+   
R+   cycles = 10,
R+   
R+   cost   = cost_total,
R+   effect = qaly,
R+   
R+   method = "life-table")
\end{Sinput}
\begin{Soutput}
base: detected use of 'state_time', expanding states: pre, symp.
\end{Soutput}
\begin{Soutput}
Fetching mortality data from package cached data.
\end{Soutput}
\begin{Soutput}
Using cached data from year 2015.
\end{Soutput}
\begin{Soutput}
Fetching mortality data from package cached data.
\end{Soutput}
\begin{Soutput}
Using cached data from year 2015.
\end{Soutput}
\begin{Soutput}
med: detected use of 'state_time', expanding states: pre, symp.
\end{Soutput}
\begin{Soutput}
surg: detected use of 'state_time', expanding states: pre, symp.
\end{Soutput}
\end{Schunk}
Strategy names are defined at that point by using the argument names provided by the user.\footnote{Strategy names are used in the results, and by functions such as \code{dispatch\_strategy()}.} We define \code{cost_total} and \code{qaly} as the respective cost and effectiveness result. The model is run for 10 cycles (i.e. 10 years), and state membership counts are corrected using the life-table method~\citep{barendregt2009}. By default the starting population is made of 1,000 patients in the first state, and no patient in the other states.

\newpage

\subsection{Results interpretation} \label{sec:run_mod} 

\emph{How do the strategies compare to each other with regard to their relative cost and effectiveness?}

The answer is given by calculating the total expected cost and effectiveness of all strategies, and then computing the incremental cost-effectiveness ratio (ICER) between them~\citep{drummond2005}. The ICER between strategies $A$ and $B$ is defined as:
$$
\frac{C_B - C_A}{E_B - E_A}
$$
Where $C$ is the total expected cost of a strategy and $E$ its total expected effect (e.g. sum of life-years of the population). Thus the ICER is the cost of an incremental unit of effectiveness. The most cost-effective strategy is (1) the most effective strategy (2) among the strategies having an ICER no higher than a threshold. This ICER threshold is the maximal willingness to pay for an additional unit of effectiveness: it is a political choice that depends on multiple factors~\citep{claxton2015}.

\begin{figure}
\centering
\begin{Schunk}

\includegraphics[width=0.5\linewidth]{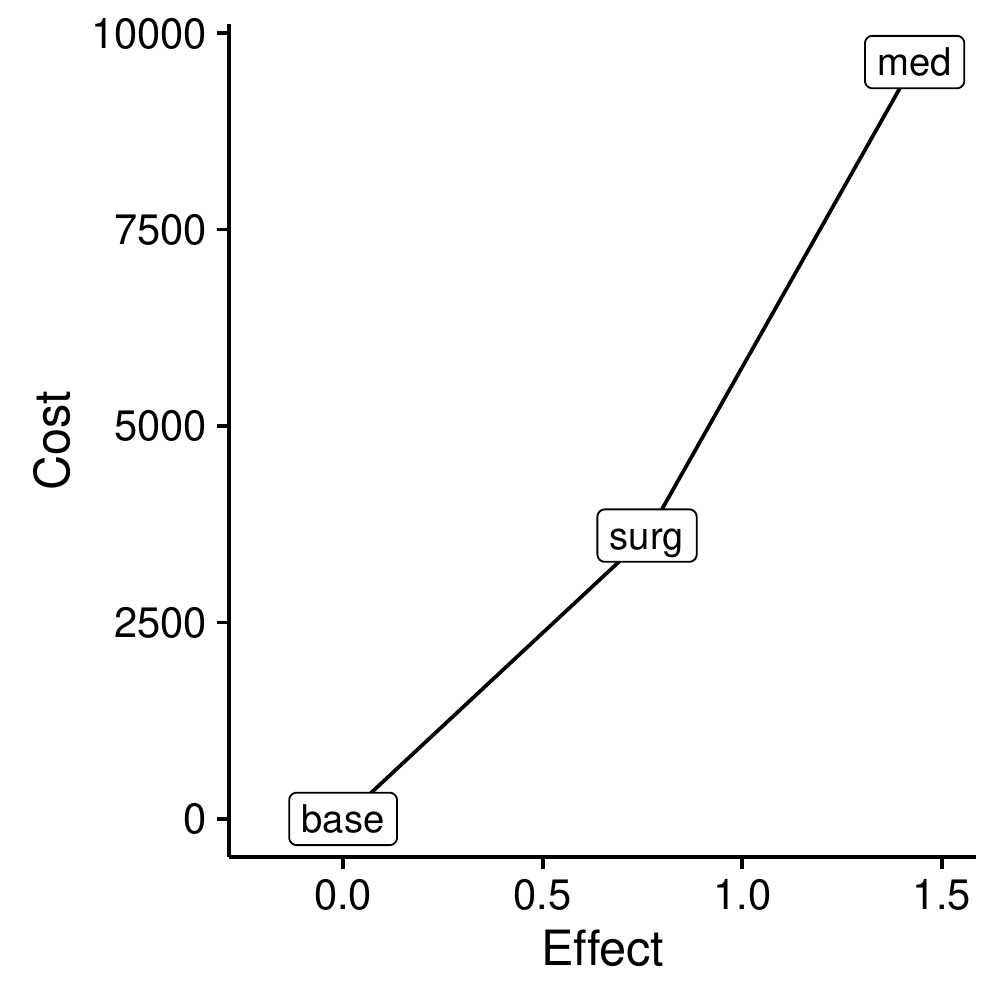} \end{Schunk}
\caption{Incremental cost and effect of strategies on the cost-effectiveness plane, taking the \code{base} strategy as a reference.}
\label{fig:res_ce}
\end{figure}

The strategies can be presented on a cost-effectiveness plane (Figure~\ref{fig:res_ce}),\footnote{Generated by \code{plot(res\_mod, type = "ce")}.} were we see that both \code{med} and \code{surg} are more effective than \code{base}, but more costly.
\begin{Schunk}
\begin{Sinput}
R> summary(res_mod, threshold = c(1000, 5000, 15000))
\end{Sinput}
\begin{Soutput}
3 strategies run for 10 cycles.

Initial state counts:

pre = 1000L
symp = 0L
death = 0L

Counting method: 'life-table'.

Values:

     cost_treat cost_hospit cost_total     qaly
base          0    54214446   42615142 5792.258
med    27619456    37181168   52246211 7224.085
surg   10074777    47429243   46220058 6553.701

Net monetary benefit difference:

      1000     5000     15000
1 8199.241 2471.932     0.000
2 5355.769 2674.230  7816.725
3    0.000    0.000 11846.341

Efficiency frontier:

base -> surg -> med

Differences:

     Cost Diff. Effect Diff.     ICER Ref.
surg   3604.915    0.7614427 4734.322 base
med    6026.153    0.6703846 8989.098 surg
\end{Soutput}
\end{Schunk}
From the printed model output presented above we see in the \code{ICER} column of the \code{Differences} section that \code{surg} is more cost-effective than \code{base} if one is willing to pay
4,734
more per QALY gained. Furthermore \code{med} is more cost-effective than \code{surg} if one is willing to pay
8,989
more per QALY gained.

A net monetary benefit analysis~\citep{stinnett1998} is run by specifying thresholds ICER values in the \code{summary()} function with the \code{threshold} argument. We see in the \code{Net monetary benefit} section that at a threshold ICER of 1,000 the strategy with the highest net monetary benefit is \code{base}, at 5,000 \code{surg} and at 15,000 \code{med}.

\begin{figure}[p]
\centering
\begin{Schunk}

\includegraphics[width=0.8\linewidth]{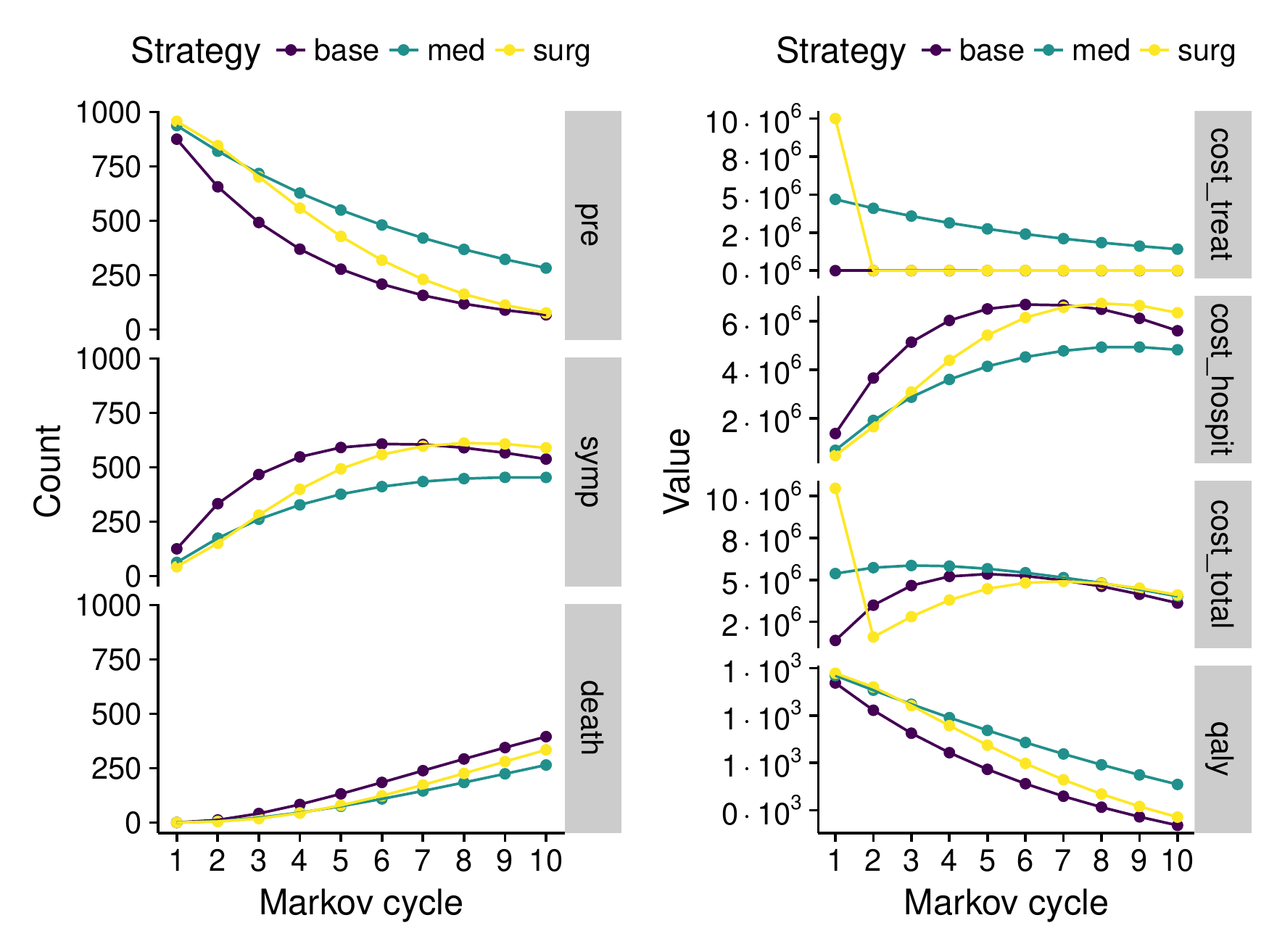} \end{Schunk}
\caption{Evolution over time of counts by state (A) and of values (B).}
\label{fig:res_count}
\end{figure}
Figure~\ref{fig:res_count} gives us more information about what happens in our model: the effect of surgery seems to wear down with time compared to the medical treatment.\footnote{Generated by \code{plot(res\_mod, type = "counts", panel = "by\_state")} and \code{plot(res\_mod, type = "values", panel = "by\_value")}.} Surgery delays the outcome, reporting degraded health status and hospital costs further in time. After a few years the hospital costs of the surgery strategy reach similar levels to the base strategy. Nevertheless these increased hospital costs do not outweigh the important treatment costs associated with the medical therapy.

\subsection{Uncertainty analysis} \label{sec:psa}

\emph{What is the uncertainty of these results? What strategy is probably the most cost-effective?}

Uncertainty of the results originate from uncertainty regarding the true value of the input parameters (e.g. treatment effect, cost of hospital stays, quality of life with the disease, survival probabilities). The effect of this uncertainty can be assessed by varying the parameter values and computing the model results with these new inputs. While multiple methods exist to study uncertainty~\citep{briggs1994}, deterministic and probabilistic sensitivity analysis (DSA and PSA) are the most widely used~\citep{briggs2006}. 

In a DSA, parameter values are changed one by one, usually to a low and high value (e.g. the lower and upper bounds of the parameter confidence interval). Model results are plotted on a \emph{tornado plot} to display how a change in the value of one parameter impacts the model results. A DSA gives a good sense of the relative impact of each parameter on the uncertainty of the model outcomes, but does not account for the total uncertainty over all the parameters, for skewed or complex parameter distribution, nor for correlations between the errors of different parameter estimates~\citep{briggs1994}.

\begin{figure}[p]
\centering
\begin{Schunk}

\includegraphics[width=0.5\linewidth]{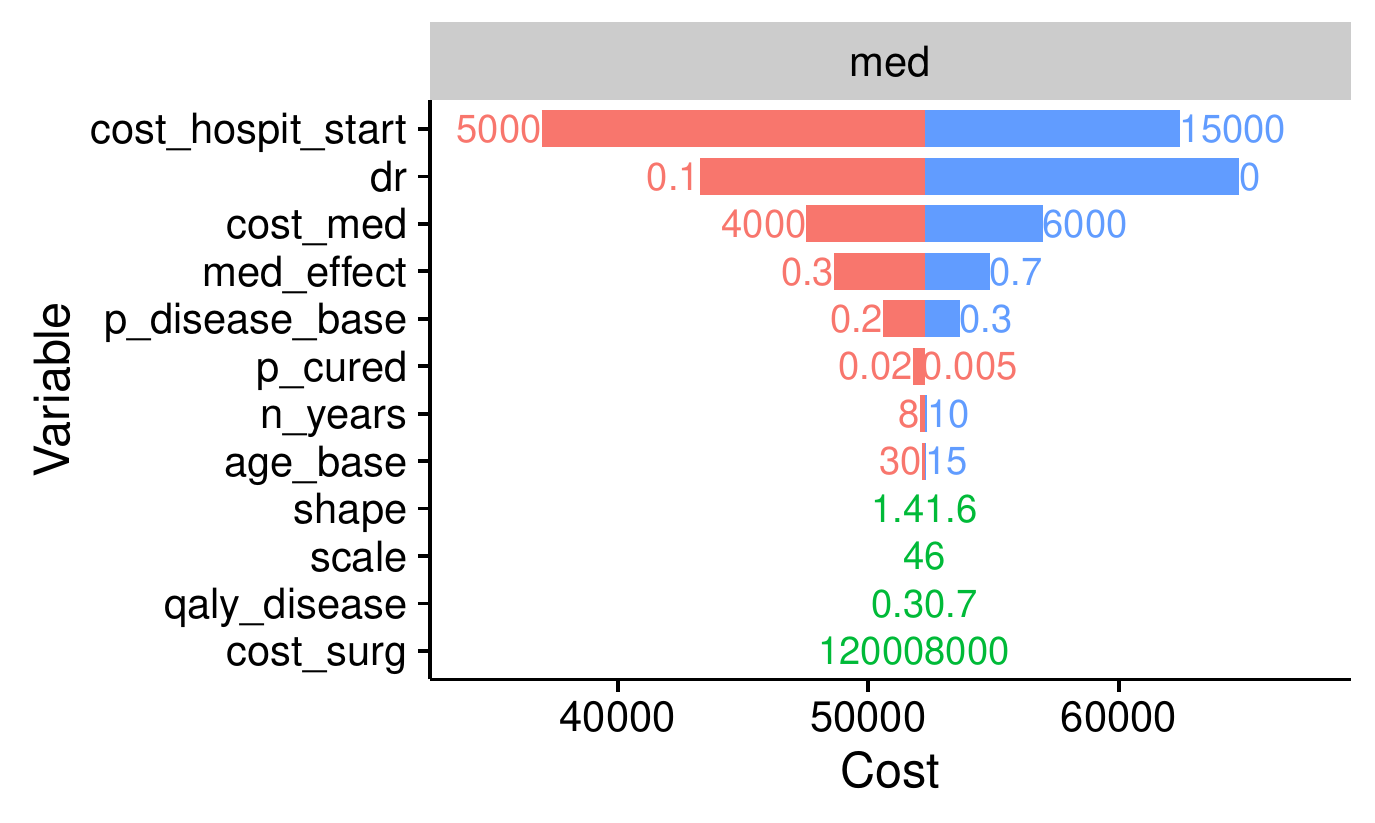} \end{Schunk}
\caption{Tornado plot presenting uncertainty of the cost for the \code{med} strategy. Each line shows how setting the parameter to its low and high value impacts costs. The bars are centred around baseline costs.}
\label{fig:dsa}
\end{figure}

We define the DSA with \code{define_dsa()} by specifying a lower and upper bound for each parameter of interest.

\begin{Schunk}
\begin{Sinput}
R> def_dsa <- define_dsa(
R+   age_base,          15,    30,
R+   p_disease_base,    0.2,   0.3,
R+   p_cured,           0.005, 0.02,
R+   med_effect,        0.3,   0.7,
R+   shape,             1.4,   1.6,
R+   scale,             4,     6,
R+   cost_med,          4000,  6000,
R+   cost_surg,         8000,  12000,
R+   cost_hospit_start, 5000,  15000,
R+   dr,                0,     0.1,
R+   qaly_disease,      0.3,   0.7,
R+   n_years,           8,     10)
\end{Sinput}
\end{Schunk}
\begin{Schunk}
\begin{Sinput}
R> res_dsa <- run_dsa(res_mod, dsa = def_dsa)
\end{Sinput}
\begin{Soutput}
Running DSA on strategy 'base'...
\end{Soutput}
\begin{Soutput}
Running DSA on strategy 'med'...
\end{Soutput}
\begin{Soutput}
Running DSA on strategy 'surg'...
\end{Soutput}
\end{Schunk}

Only parameters (e.g. state values, transition probabilities) defined with \code{define_parameters()} can be modified in a DSA (or a PSA).  Accordingly, many state values, transition probabilities, and the shape and scale parameters used in our example were defined as parameters, thus allowing them to be varied in sensitivity analyses. Once defined, the analysis can be run using \code{run_dsa()}.

Figure~\ref{fig:dsa} shows the impact of varying each parameter individually on total cost for the \code{med} strategy.\footnote{Generated by \code{plot(res\_dsa, result = "cost", strategy = "med")}.} The results demonstrate that the discount rate and hospital costs have a greater impact than other parameters. Unsurprisingly parameters used only in the surgery strategy and parameters unrelated to costs have no effect on the cost of the \code{med} strategy.

In a PSA, the model is re-run for a given number of simulations with each parameter being replaced with a value re-sampled from a user-defined probability distribution. These results are then aggregated, allowing us to obtain the probability distribution of model outputs~\citep{critchfield1986}.

\begin{figure}
\centering
\begin{Schunk}

\includegraphics[width=\maxwidth]{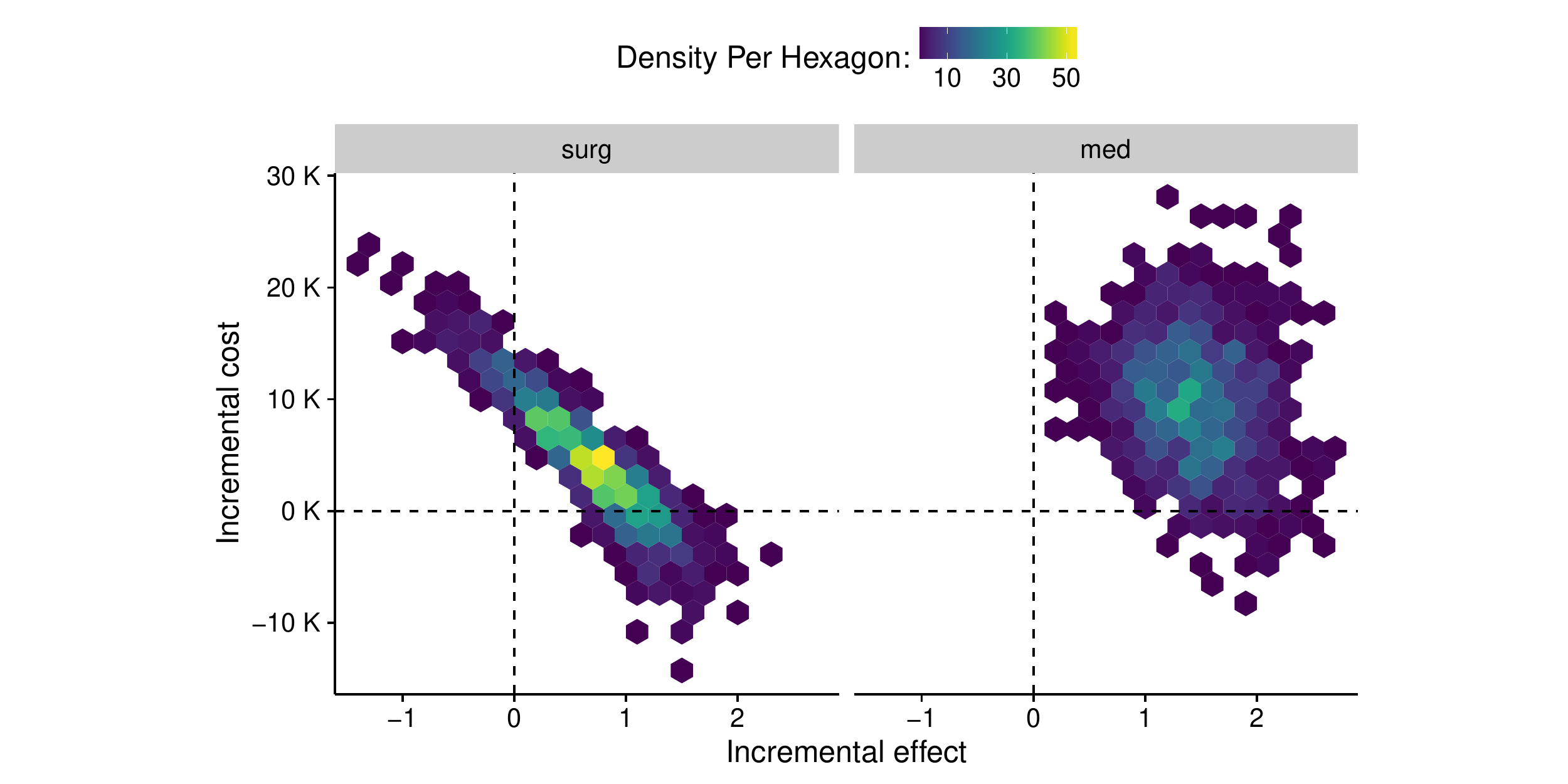} \end{Schunk}
\caption{Uncertainty of the incremental cost and effect of strategies on the cost-effectiveness plane, taking the \code{base} strategy as a reference.}
\label{fig:ce_psa}
\end{figure}

We define the parameter distributions with \code{define_psa()}, and optionally their correlation structure with \code{define_correlation()}. 

\begin{Schunk}
\begin{Sinput}
R> def_psa <- define_psa(
R+   age_base          ~ normal(mean = 20, sd = 5),
R+   p_disease_base    ~ binomial(prob = 0.25, size = 500),
R+   p_cured           ~ binomial(prob = 0.001, size = 500),
R+   med_effect        ~ lognormal(mean = 0.5, sd = 0.1),
R+   shape             ~ normal(mean = 1.5, sd = 0.2),
R+   scale             ~ normal(mean = 5, sd = 1),
R+   cost_med          ~ gamma(mean = 5000, sd = 1000),
R+   cost_surg         ~ gamma(mean = 20000, sd = 3000),
R+   cost_hospit_start ~ gamma(mean = 11000, sd = 2000),
R+   dr                ~ binomial(prob = 0.05, size = 100),
R+   qaly_disease      ~ normal(mean = 0.5, sd = 0.1),
R+   n_years           ~ poisson(mean = 9),
R+   
R+   correlation = define_correlation(
R+     shape,    scale,          -0.5,
R+     age_base, p_disease_base,  0.3))
\end{Sinput}
\end{Schunk}
\begin{Schunk}
\begin{Sinput}
R> res_psa <- run_psa(res_mod, psa = def_psa, N = 1000)
\end{Sinput}
\begin{Soutput}
Resampling strategy 'base'...
\end{Soutput}
\begin{Soutput}
Resampling strategy 'med'...
\end{Soutput}
\begin{Soutput}
Resampling strategy 'surg'...
\end{Soutput}
\end{Schunk}
We then run the PSA with \code{run_psa()}, here for 1,000 re-samplings. The results can be plotted as uncertainty clouds on the cost-effectiveness plane (Figure~\ref{fig:ce_psa}).\footnote{A similar plot can be generated with \code{plot(res\_psa, type = "ce")}.}

\begin{figure}
\centering
\begin{Schunk}

\includegraphics[width=\maxwidth]{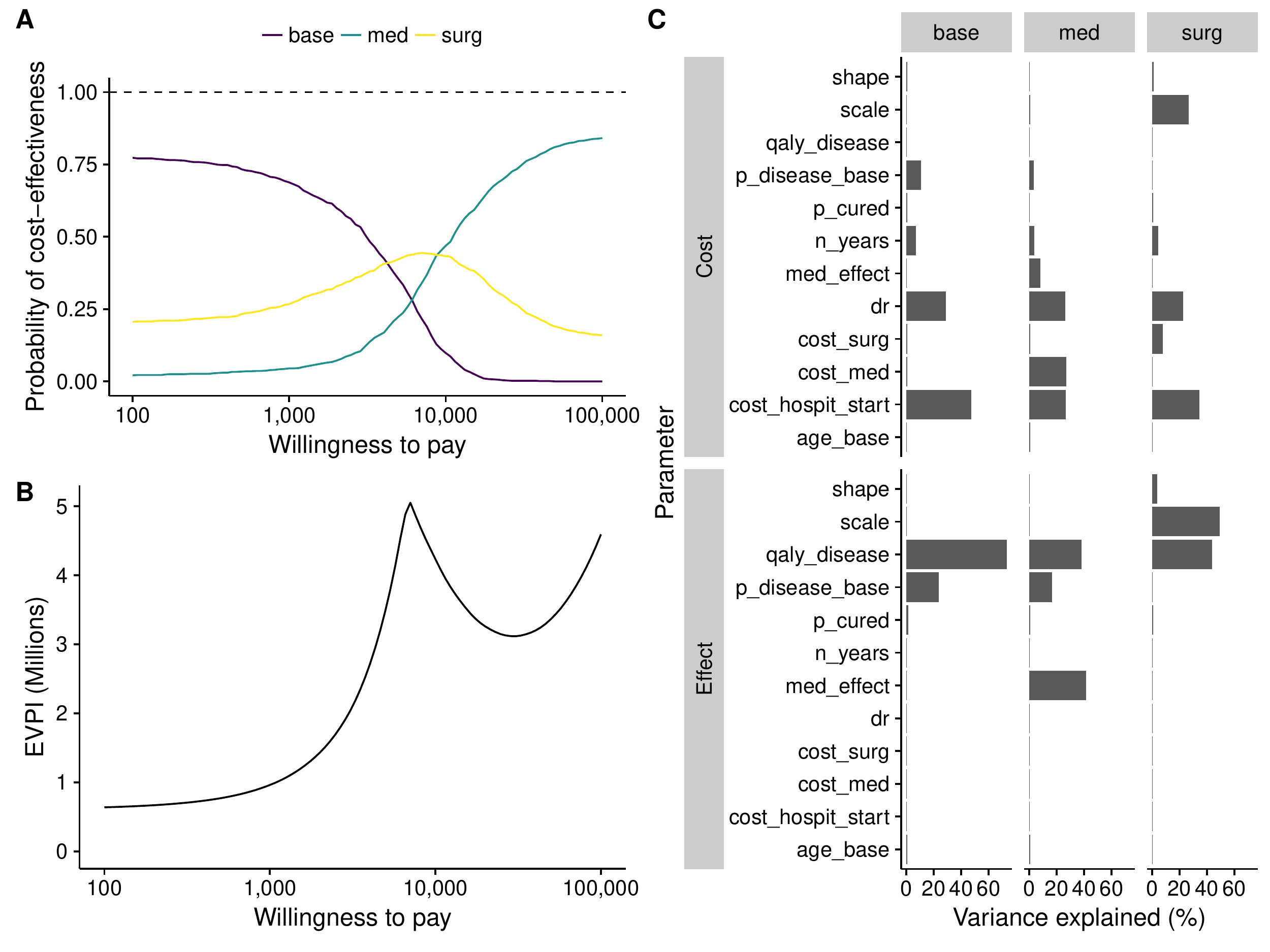} \end{Schunk}
\caption{A: Cost-effectiveness acceptability curve; B: Expected value of perfect information; C: Covariance analysis of PSA results.}
\label{fig:ceac_psa}
\end{figure}

The probability of a strategy being cost effective can be plotted for various willingness to pay values on a cost-effectiveness acceptability curve~\citep{vanhout1994,fenwick2005}. In Figure~\ref{fig:ceac_psa}A we see that with a threshold ICER below 1,000 the \code{base} strategy is probably the most cost-effective.\footnote{Generated by \code{plot(res\_psa, type = "ac")}.} Above 50,000, \code{med} is probably the most cost-effective strategy. Between those 2 values the decision is less clear.

It is possible to compute the expected value of perfect information (EVPI) depending on the willingness to pay~\citep{claxton1996,felli1997}. This is a quantification of the cost of potentially choosing the wrong strategy, and thus conversely the price one is ready to pay to reduce the risk of incorrect decisions by obtaining more information (e.g. by conducting more studies). In Figure~\ref{fig:ceac_psa}B we see the EVPI peaks between 1,000 and 10,000, where the uncertainty is high.\footnote{Generated by \code{plot(res\_psa, type = "evpi")}.} It also increases for higher willingness to pay, because even though the uncertainty is not as high, the costs of a wrong decision become higher.

The EVPI can indicate whether conducting more research is cost-effective. But it does not inform on the value of getting more information on particular parameters~\citep{briggs2006}. The expected value of perfect information for parameters (EVPPI) is very similar to the EVPI, but returns values by parameters~\citep{ades2004}. Unfortunately its computation is not trivial. PSA results can be exported to compute EVPPI with the Sheffield Accelerated Value of Information \proglang{SAVI} software~\citep{savi} with \code{export\_savi()}.

The individual contribution of parameter uncertainty on the overall uncertainty~\citep{briggs2006} is illustrated by Figure~\ref{fig:ceac_psa}C.\footnote{Generated by \code{plot(res\_psa, type = "cov")}. We could also perform the same analysis on the difference between strategies with the option \code{diff = TRUE}.} We can see that, depending on the strategy, different parameters generate the uncertainty on costs and effect. In all cases \code{dr}, \code{cost_hospit_start} and \code{qaly_disease} explain a high part of variability for all strategies. Unsurprisingly the effect of \code{scale} (the scale of the post-surgery Weibull survival function), \code{med_effect} and \code{cost_med} are limited to the \code{surg} or \code{med} strategies.

In addition, average model values can be computed on the results and presented in a summary similar to the \code{run_model()} output. Because of non-linearities in Markov models, averages over the PSA output distribution are more accurate than point estimates~\citep{briggs2006}. In our case the ICERs changed from 4,734 to 6,453
and 8,989 to 7,059 for the \code{surg} and \code{med} strategies respectively.

\subsection{Heterogeneity analysis}

\emph{How does the cost-effectiveness of strategies vary depending on the characteristics of the population?}

If population characteristics are available, model results can be computed on the different sub-populations to study the heterogeneity of the resulting model outputs~\citep{briggs2006}. Furthermore, average population-level results can be computed from these distributions.

\begin{figure}
\centering
\begin{Schunk}

\includegraphics[width=0.5\linewidth]{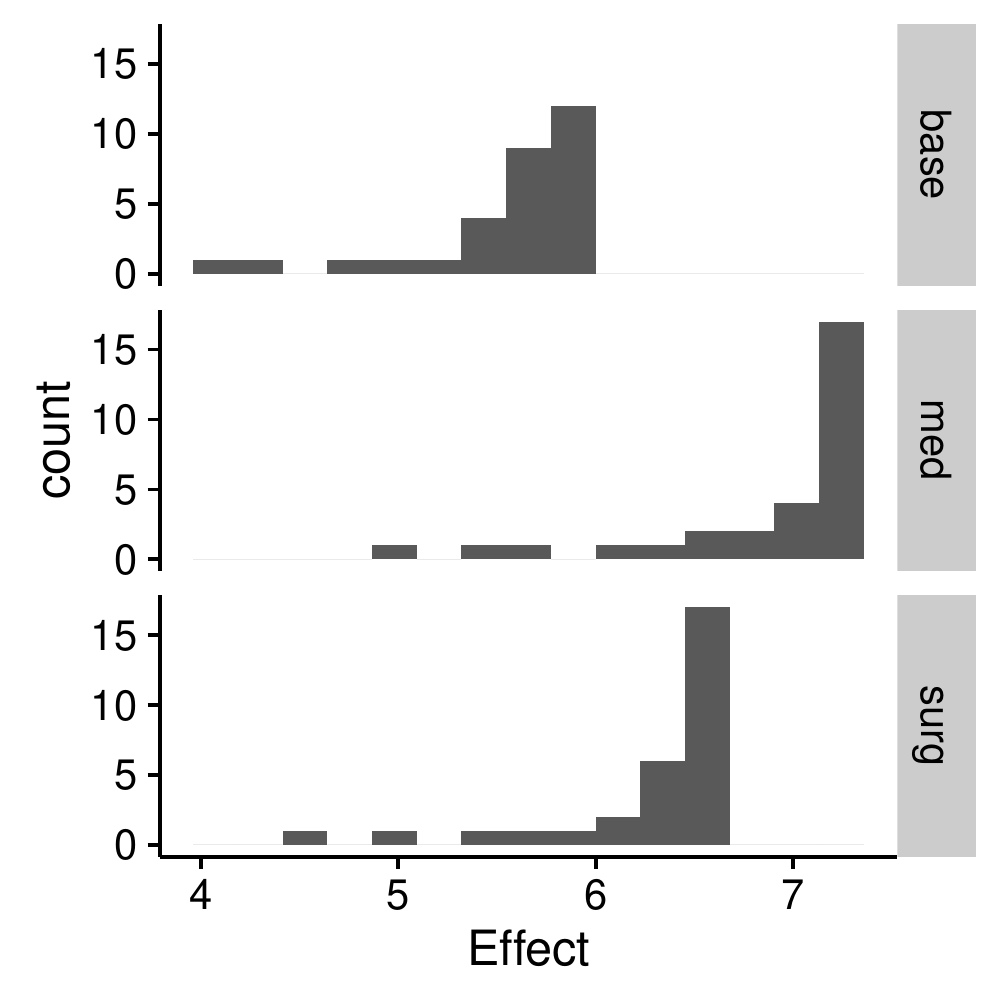} \end{Schunk}
\caption{Heterogeneity of strategy effects in the population.}
\label{fig:update}
\end{figure}
The model we ran in Section~\ref{sec:run_mod} computed results for a cohort of males aged 20. To assess how population characteristics affect model results we can run a heterogeneity analysis. We use the \code{update()} function to run the model on a table containing population data.\footnote{In this example we use a table with population characteristics, here named \code{tab\_pop}, with an optional column \code{.weights} giving the relative population weight of each strata.}

\begin{Schunk}
\begin{Sinput}
R> head(tab_pop)
\end{Sinput}
\begin{Soutput}
  age_base sex_indiv   .weights
1       10       MLE 0.04242889
2       10      FMLE 0.86696571
3       15       MLE 0.69960873
4       15      FMLE 0.51253057
5       20       MLE 0.91723545
6       20      FMLE 0.09685623
\end{Soutput}
\end{Schunk}
\begin{Schunk}
\begin{Sinput}
R> pop_mod <- update(res_mod, newdata = tab_pop)
\end{Sinput}
\begin{Soutput}
Updating strategy 'base'...
\end{Soutput}
\begin{Soutput}
Updating strategy 'med'...
\end{Soutput}
\begin{Soutput}
Updating strategy 'surg'...
\end{Soutput}
\end{Schunk}
The summary of the updated model gives the distribution of the values of interest in the population, and the average model values over the entire population. Here the average ICERs in the population are 5,052 and 9,150 for the \code{surg} and \code{med} strategies respectively, quite similar to the values computed in Section~\ref{sec:run_mod} (4,734 and 8,989). We can also plot the distribution of model results (e.g. the intervention effects in Figure~\ref{fig:update}).\footnote{Generated by \code{plot(pop\_mod, result = "effect", bins = 15)}.}

\subsection{Budget impact analysis}

\emph{What would be the total cost of a strategy for the health system?}

So far we mostly worked on model results at the scale of the individual (e.g. cost per person). If we want to implement a strategy at a health system level we also need to know the total cost over a given time horizon, in order to assess whether the strategy is sustainable. This is called a budget impact analysis (BIA). The main differences with the classic model are (1) the patient counts at the model start should reflect the population statistics, and (2) additional patients may enter the model every year (new disease cases).

We use the \code{init} and \code{inflow} arguments of \code{run_model()} to implement BIA, here for the \code{med} strategy. The inflow of new patients is defined with \code{define_inflow()}. Inflow counts can depend on \emph{model time} (\emph{state time} dependency is meaningless in this context).
\begin{Schunk}
\begin{Sinput}
R> res_bia <- run_model(
R+   parameters = par_mod,
R+   
R+   med = strat_med,
R+   
R+   cycles = 10,
R+   
R+   cost = cost_total,
R+   effect = qaly,
R+   
R+   method = "life-table",
R+   
R+   init =   c(
R+     pre   = 25000,
R+     symp  = 5000,
R+     death = 0),
R+   inflow = define_inflow(
R+     pre   = 8000,
R+     symp  = 0,
R+     death = 0))
\end{Sinput}
\begin{Soutput}
med: detected use of 'state_time', expanding states: pre, symp.
\end{Soutput}
\end{Schunk}
At the start of the model there are 25,000 patients with asymptomatic shame and 5,000 with a symptomatic form of the disease in the population. Every year 8,000 additional cases of shame are added to the model, starting the disease in the asymptomatic state.

\begin{Schunk}
\begin{Sinput}
R> summary(res_bia)
\end{Sinput}
\begin{Soutput}
1 strategy run for 10 cycles.

Initial state counts:

pre = 25000
symp = 5000
death = 0

Counting method: 'life-table'.

Values:

    cost_treat cost_hospit cost_total     qaly
med 1942366603  2621681008 3531335211 508495.2
\end{Soutput}
\end{Schunk}
The total cost of strategy \code{med} over a 10-year time horizon will be 3.5
billions.

\section{Other features and extensions} \label{sec:features}

This section introduces features and extensions that were not presented in the previous example.

\subsection{Survival analysis}

The \pkg{heemod} package provides a number of ways to estimate transition probabilities from survival distributions. Survival distributions can come from at least three different sources:

\begin{itemize}
  \item User-definded parametric distributions created using the \code{define_survival()} function.
  \item Fitted parametric distributions with \code{flexsurv::flexsurvreg()}~\citep{flexsurv}.
  \item Fitted Kaplan-Meiers with \code{survival::survfit()}~\citep{survival-package,survival-book}.
\end{itemize}

Once defined, each of these types of distributions can be combined and modified using a standard set of operations. Treatment effects can be applied to any survival distribution:

\begin{itemize}
  \item Hazard ratio: \code{apply_hr()}.
  \item Odds ratio: \code{apply_or()}.
  \item Acceleration factor: \code{apply_af()}.
\end{itemize}

In addition, distributions can be combined:

\begin{itemize}
  \item Join one (or more) survival distributions together: \code{join()}.
  \item Pool two (or more) survival distributions: \code{pool()}.
  \item Combine two (or more) survival distributions as independent risks: \code{add_hazards()}.
\end{itemize}

The transition or survival probabilities are computed with \code{compute_surv()}. Time (usually \code{model_time} or \code{state_time}) needs to be passed to the function as a \code{time} argument.

All these operations can be chained with the \code{\%>\%} piping operator \citep{magrittr}, e.g.:
\begin{Schunk}
\begin{Sinput}
R> fit_cov %>% 
R+   apply_hr(hr = 2) %>% 
R+   join(
R+     fitcov_poor,
R+     at = 3) %>%
R+   pool(
R+     fitcov_medium,
R+     weights = c(0.25, 0.75)) %>%
R+   add_hazards(
R+     fit_w) %>% 
R+   compute_surv(time = 1:5)
\end{Sinput}
\end{Schunk}

\subsection{Convenience functions}

For reproducibility and ease of use we implemented convenience functions to perform some of the most common calculations needed in health economic evaluation studies (e.g converting incidence rates, odds ratios, or relative risks to transition probabilities with \code{rate_to_prob()}, \code{or_to_prob()}, or \code{rr_to_prob()} respectively). Probabilities and discount rates can be rescaled to fit different time frames (generally the duration of a cycle) with \code{rescale_prob()} and \code{rescale_discount_rate()}.

\subsection{Cluster computing}

PSA and heterogeneity analyses can become time-consuming since they consist in iteratively re-running the model with new parameter inputs. Because this workload is \emph{embarrassingly parallel}~\citep{herlihy}, i.e. there is no dependency or need for communication between the parallel tasks, it can easily be run on a cluster relying on the \pkg{parallel}~\citep{r} package. This is done by calling the \code{use_cluster()} function. This function can either take as an argument:
\begin{enumerate}
  \item A number: a local cluster with the given number of cores will be created.
  \item A cluster object defined with the \code{makeCluster()} function from \pkg{parallel}: the user-defined cluster will be used (e.g. to use more complex clusters with non-local hosts).
\end{enumerate}

\subsection{Alternative interfaces}

To facilitate the use of \pkg{heemod} by users not familiar with \proglang{R} we developed a \pkg{shiny}~\citep{shiny} graphical user interface. Similarly, for users that require the use of spreadsheet models (such as health regulatory agencies), a model can be specified in spreadsheet files and run by \pkg{heemod}. To keep the traceability, transparency and reproducibility advantages provided by written source code it is possible to export models built from these interfaces to \proglang{R} source code files.

These alternative interfaces are needed in a context where (1) Markov models are already widely used and implemented on spreadsheet software, (2) a significant proportion of the modellers are not \proglang{R} users, and (3) health regulatory agencies from several countries require models to be in spreadsheet format. We believe that to gain acceptance in a field such as health economic evaluation where habits are already ingrained one must adapt to existing user requirements, as long as the final outcome is to help develop transparency and reproducibility in the domain.

\subsection{Extension to other types of model}

Even though the main focus of \pkg{heemod} is to compute Markov models, other methods that model state changes can be included in the package: only the \code{transition} argument of \code{define_strategy()} and the associated evaluation methods need to be extended.

For example partitioned survival models~\citep{williams2016_part} were added to the package recently. These models can be computed by passing an object defined by \code{define_part_surv()} to \code{transition}.

In theory most modelling methods that return state counts over time could be integrated into \pkg{heemod}, e.g. dynamic models for infectious diseases~\citep{keeling2011,snedecor2012}.

\section{Mathematical implementation} \label{sec:math}

In this section we detail the mathematical implementation of most of the features of \pkg{heemod}.

\subsection{Parameter correlation in PSA}

Correlation of parameters in PSA was implemented with the following steps:
\begin{enumerate}
  \item A correlation structure is define with \code{define_correlation()} (see Section~\ref{sec:psa}).
  \item Values are sampled from a multi-normal distribution having the required correlation structure with \pkg{mvnfast}~\citep{mvnfast}.
  \item The sampled values are then mapped to the target distributions on a quantile by quantile basis.
\end{enumerate}

This approach described in \citet{briggs2006} is an approximation that allows to define correlations between arbitrary distributions. The final Pearson correlation coefficients between the target distributions may differ slightly from the ones initially defined by the user. That issue is mostly true if the target distributions are too dissimilar, e.g. a gamma and a binomial distribution.

\subsection{Time-dependency implementation}

A throughout description of time-dependency in Markov models is given by \citet{hawkins2005}, with solutions for the computation of both non-homogeneous and semi-Markov models. These methods were implemented in the \pkg{heemod} package. Markov models with \emph{model time} dependency are usually called non-homogeneous Markov models, and models with \emph{state time} dependency are called semi-Markov models.

\emph{Model time} (non-homogeneous Markov models) was implemented by using a 3-dimensional transition matrix $U$. As in the 2-dimensional matrix $T$ described in Section~\ref{sec:intro} the indices of the first 2 dimensions $i,j$ encode the transition probability between state $i$ and $j$. In addition the third dimension index $k$ corresponds to the number of cycles the model has run so far, so that element $i,j,k$ of matrix $U$ corresponds to the transition probability between state $i$ and $j$ at time $k$. The probability of being in a given state at time $t$ is given by a simple extension of Equation~\ref{eq:mat_2}:
\begin{equation} \label{eq:mat_3}
X \times \prod_{k=1}^{t}U_{k}
\end{equation}
Where $X$ is a vector\footnote{Of length equal to the number of states.} giving the probability of being in a given state at the start of the model, $\prod$ stands for matrix multiplication, and $U_{k}$ is a 2-dimensional slice of the 3-dimensional transition matrix $U$, giving the transition probabilities at time $k$.

\emph{State time} (semi-Markov models) was implemented with the tunnel-state method, described in \citet{hawkins2005}. A tunnel state is a state that can be occupied for only 1 cycle, it represents at the same time the health state a person is in and the number of cycles previously spent in this state. A state $A$ with state-time dependency is expanded in $t$ tunnel states $A_{1}, A_{2}, \dots, A_{t}$ (where $t$ is the total number of cycles). For example consider the following transition matrix:
\begin{equation} \label{eq:unexp}
\begin{bmatrix}
P(A\rightarrow A)=f(s) & P(A\rightarrow B)=C \\
P(B\rightarrow A) & P(B\rightarrow B)
\end{bmatrix}
\end{equation}
Where $P(A\rightarrow B)$ is the transition probability between state $A$ and $B$, $s$ the number of cycles spent in state $A$, $f$ an arbitrary function returning a transition probability, and $C$ the probability complement (1 minus the sum of probabilities in a given row). $P(B\rightarrow A)$ and $P(B\rightarrow B)$ are arbitrary probabilities that do not depend on state time.

The matrix in Equation~\ref{eq:unexp} can be expanded to the following matrix when the model is run for $t$ cycles:
\begin{equation} \label{eq:exp}
\left[
\begin{smallmatrix}
0      & P(A_{1}\rightarrow A_{2})=f(1)   & 0      & \cdots & 0 & 0      & P(A_{1}\rightarrow B)=C      \\
0      & 0      & P(A_{2}\rightarrow A_{3})=f(2)   & \cdots& 0  & 0      & P(A_{2}\rightarrow B)=C      \\
0      & 0      & 0      & \cdots & 0 & 0      & P(A_{3}\rightarrow B)=C      \\
\vdots & \vdots & \vdots & \ddots & \vdots & \vdots & \vdots \\
0      & 0      & 0      & \cdots & 0 & P(A_{t-1}\rightarrow A_{t})=f(t-1)   & P(A_{t-1}\rightarrow B)=C      \\
0      & 0      & 0      & \cdots & 0 & P(A_{t}\rightarrow A_{t})=f(t)      & P(A_{t}\rightarrow B)=C      \\
P(B\rightarrow A) & 0      & 0      & \cdots & 0 & 0      & P(B\rightarrow B) \\
\end{smallmatrix}
\right]
\end{equation}
The semi-Markov model described in Equation~\ref{eq:unexp} is now rearranged as a classic Markov model. It can be noticed that if we were to run this model for more than $t$ cycles then $P(A\rightarrow A)$ would remain constant after time $t$, at a value of $f(t)$. This property is useful in situations where $f(s)$ become roughly constant when $s \geq t$: in that case we can stop the state expansion at $t$ tunnel states in order to limit the final matrix size, and thus the computational burden. This approximation is implemented in \pkg{heemod} with the \code{state_cycle_limit} option of \code{run_model()}.

In practice in \pkg{heemod} any state where \emph{state time} dependency is detected is implicitly converted internally to a sequence of tunnel states. Counts and values are internally computed for each tunnel state, and then internally re-aggregated before being returned to the user as a single state. The transformation to a sequence of tunnel states is thus invisible for the user, except for a message informing of the implicit state expansion.

\subsection{Implementing budget impact analysis}

The main technical difficulty to implement budget impact analyses is that new individuals may enter the model at any point in time. The classic Markov model computation described in Equation~\ref{eq:mat_2} cannot be used any more. Instead the probability of being in a given state at time $k$ is given by the following sequence:
\begin{equation} \label{eq:bia_1}
\begin{split}
a_{0} &=  Y \\ 
a_{k} &=  a_{k-1} \times T + Z
\end{split}
\end{equation}
Where $Y$ is a vector of the number of individuals in a given state at the start of the model, $T$ a 2-dimensional transition matrix, and $Z$ a vector\footnote{Of length equal to the number of states.} giving the number of new individuals entering the model at each new cycle.

Equation~\ref{eq:bia_1} can be adapted to allow (1) for \emph{model time} dependency of transition probabilities as described in Equation~\ref{eq:mat_3} and (2) for time-dependency of the number of individual entering the model at each new cycle, in this way:
\begin{equation} \label{eq:bia_2}
\begin{split}
a_{0} &=  Y \\ 
a_{k} &=  a_{k-1} \times U_{k} + Z_{k}
\end{split}
\end{equation}
Where $U_{k}$ is a 2-dimensional slice of the 3-dimensional transition matrix $U$, and $Z_{k}$ a vector giving the number of new individuals entering the model at cycle $k$.

Finally, \emph{state time} dependency by tunnel state expansion can be integrated without any change to Equation~\ref{eq:bia_2}.

\section{Package design and back-end} \label{sec:impl}

In this section we explain the ideas underlying the design of the package, the general structure, and the validation process. The entire workflow is summarised in a chart presented in Figure~\ref{fig:flow}. We then present the package back-end and how most of the features were actually implemented.

\subsection{Package design and workflow}

The package was focused on reproducibility of analyses and ease of use. Both those objectives could be reached by making the functions unambiguous and easily readable by humans. We tried to rely on Hadley Wickham's \emph{tidy manifesto} to design our package in that direction~\citep{tidyverse}.

\begin{figure}
\centering
\includegraphics[width=\textwidth]{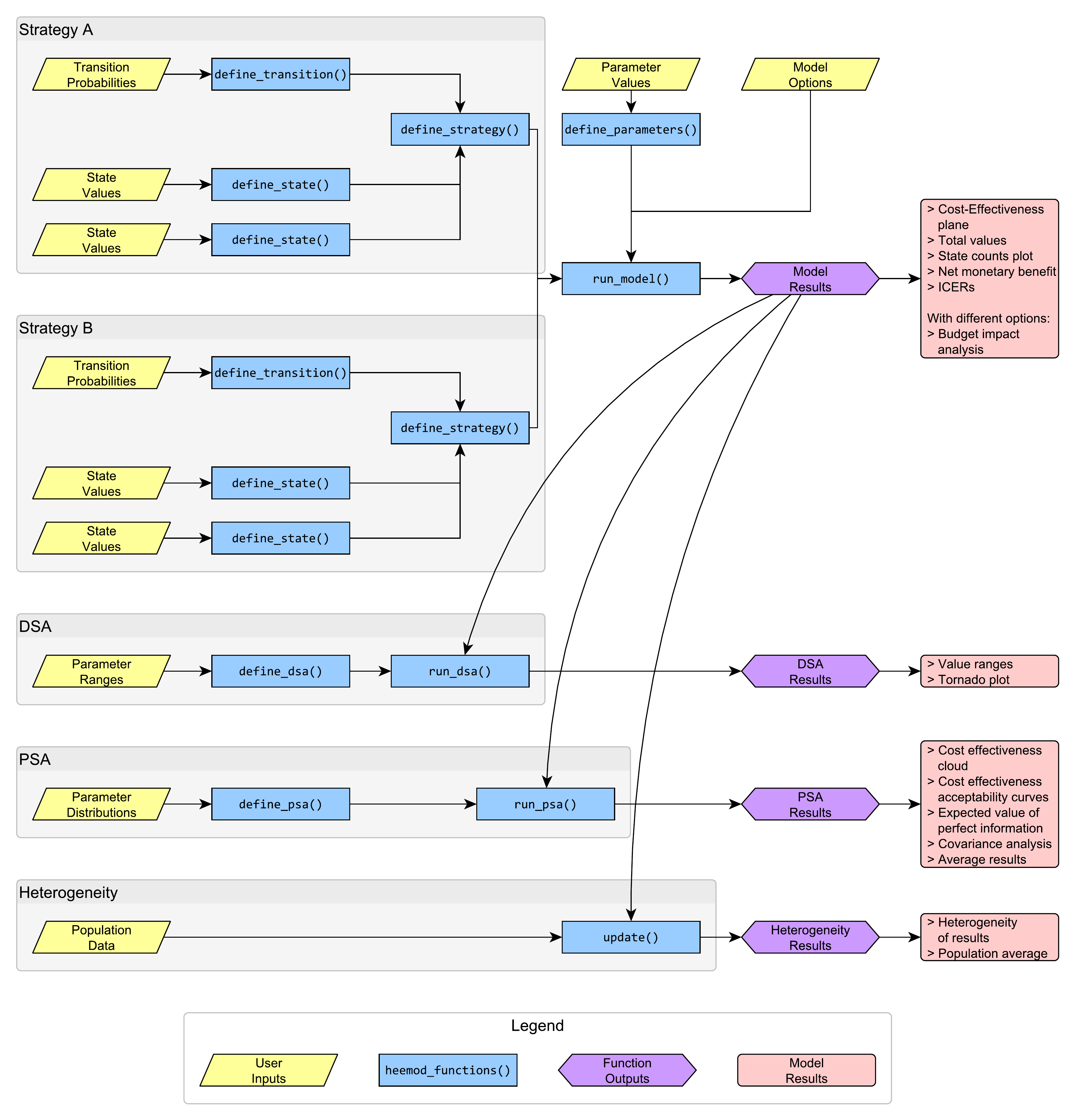}
\caption{\pkg{heemod} package workflow. Drawn with \proglang{yEd}~\citep{yed}.}
\label{fig:flow}
\end{figure}

We divided health economic evaluation modelling into distinct and sequential tasks, and wrote simple \emph{verb} functions corresponding to the most common tasks (e.g. \code{define_*}, \code{run_*}), detailed in Figure~\ref{fig:flow}.

We tried to keep functions as simple as possible: each function should do one thing, each task should have its own function. By simplifying the options we hoped to simplify how the user thinks about modelling, hence making it easier not only to build models, but more importantly for another user to read and understand someone else's model. This last point is of particular importance if we want more research transparency and reproducibility in health economic evaluation studies.

To paraphrase Hal Abelson, we think that models must be written for people to read, and only incidentally for machines to execute.

\subsection{Validation}

We validated the package by reproducing the exact result of 2 analyses described in the reference textbook by \citet{briggs2006}: the HIV therapy and the total hip replacement model. In both case we found identical results (total values, patient counts and ICERs).

To ensure the results remain correct when the package is updated or when a dependency is upgraded multiple tests were written with the \pkg{testthat} package~\citep{testthat}. We verify that functions produce the expected output, that incorrect inputs generate errors, and that results from published models are reproduced. The tests are run as soon as a modification is made to the code: when a change introduces a bug a warning is raised and remains until the issue is fixed.

\subsection{Back-end}

The \pkg{heemod} package syntax relies heavily on the non-standard evaluation features offered by \pkg{lazyeval}~\citep{lazyeval}. This allows the user to define parameters, state values, and transition probabilities as a sequence of expressions to be evaluated at runtime.  In addition to being familiar to users of the \pkg{dplyr} package~\citep{dplyr}, this results has the advantage of resembling spreadsheet formulae and making heemod more approachable, while keeping namespace collisions in check.\footnote{A usual pitfall of non-standard evaluation in \proglang{R}.}

More generally, most of \code{heemod}'s core functions rely on the \pkg{dplyr} package: the objects, data, and results are stored as \code{tbl_df} objects, the \pkg{dplyr} implementation of data frames. This is another principle of the \emph{tidy manifesto}~\citep{tidyverse}: reuse existing data structures. This allows the use of powerful base functions, efficient computation by limiting copy creation, and iterative model re-computation for PSA or heterogeneity analyses with the \code{dplyr::do()} function.

Relying on another package instead of writing package-specific functions has the drawback that slightly ill-fitting data structures may sometimes be used. But we think this drawback is outweighed by the multiple advantages of \emph{piggybacking} a widely used package such as \code{dplyr}. We benefit from the code quality control and the constantly improving features of a popular package, letting us focus our development time on actually implementing Markov models. Even more importantly our internal code is easier to understand by anyone familiar with \code{dplyr}. This last point reduces barriers to entry for potential contributors.

\begin{figure}
\centering
\begin{Schunk}

\includegraphics[width=0.8\linewidth]{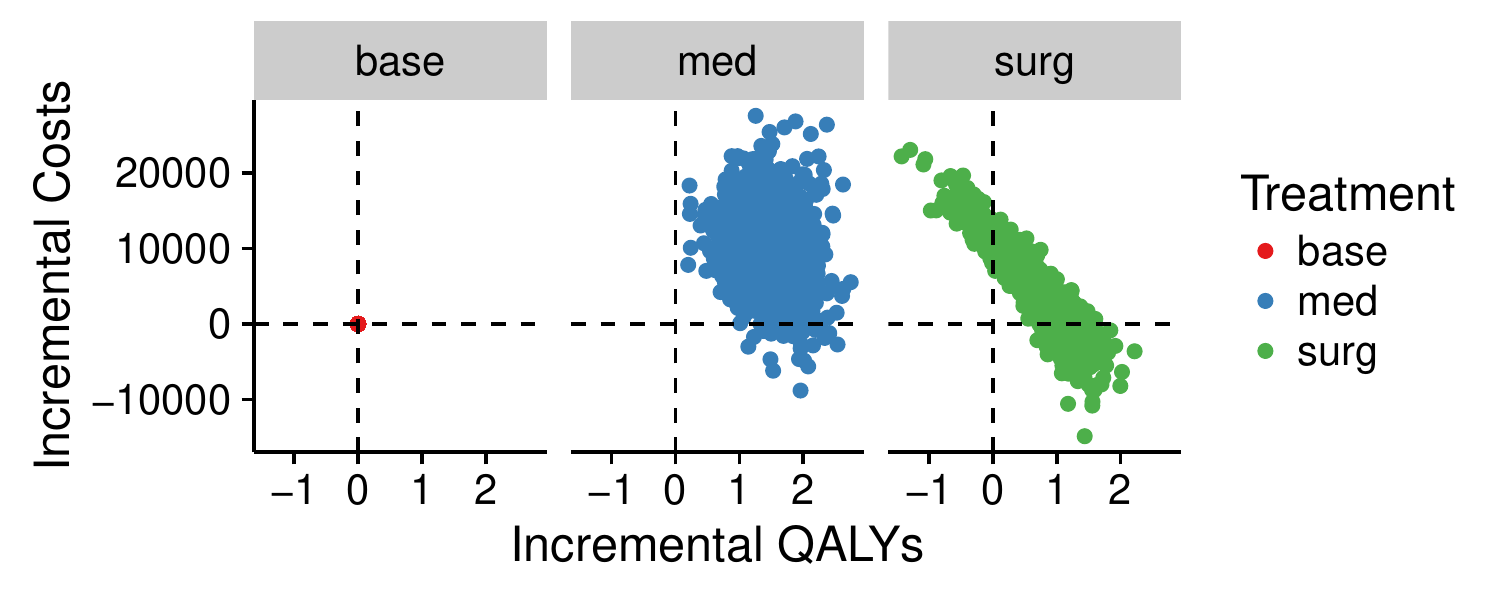} \end{Schunk}
\caption{Example of plot customization.}
\label{fig:example}
\end{figure}
The plotting functions rely on the \pkg{ggplot2} package~\citep{ggplot2}, and can be easily customized using the \code{+} operator. The following code is used to produce Figure~\ref{fig:example}:
\begin{Schunk}
\begin{Sinput}
R> library(ggplot2)
R+ 
R+ plot(res_psa, type = "ce") +
R+   scale_color_brewer(name = "Treatment", palette = "Set1") +
R+   facet_wrap(~ .strategy_names) +
R+   xlab("Incremental QALYs") + ylab("Incremental Costs") +
R+   geom_hline(yintercept = 0, linetype = "dashed") + 
R+   geom_vline(xintercept = 0, linetype = "dashed")
\end{Sinput}
\end{Schunk}
The plotting of transition matrices as directed diagrams is performed by the \pkg{diagram} package~\citep{diagram}, the covariance analysis of PSA relies on the \pkg{mgcv} package~\citep{mgcv}, and the weighted summary of the results is computed with the \pkg{Hmisc} package~\citep{hmisc}.

The running time of some functions is not negligible (e.g. \code{get_who_mr()}). While half a second is not an issue when a function is run only once, it becomes a major hurdle during re-sampling where the function may be called thousands of times. We used the memoisation features of the \pkg{memoise} package~\citep{memoise} to shorten execution time: when a function is called the result is kept in memory alongside the values of the calling arguments. If the function is called again with the same argument values then the function body is not evaluated, but the memoised result is instantly returned instead. The use of memoisation is particularly efficient in re-sampling because in most cases the values of the arguments of most functions remain identical, resulting in the same outputs.

\section*{Acknowledgements}

Thanks to Matthew Wiener, Zdenek Kabat, Vojtech Filipec and Jordan Amdahl for their contributions to this package, to Marina Filipovi\'c-Pierucci for her help with mathematical notation and for proof-reading, and to Jordan Amdahl, Corine Baayen, Yonatan Carranza, Rana Maroun, Chlo\'e Le Cossec and Guillaume Pressiat for proof-reading this paper.

The colours of the figures in this article were made with the \pkg{viridis} package~\citep{viridis}.

\bibliography{biblio}

\end{document}